\documentclass[conference,a4paper]{IEEEtran}
\IEEEoverridecommandlockouts

\usepackage{graphicx}
\usepackage[table,xcdraw]{xcolor}
\usepackage{multirow}
\usepackage{amsmath,amssymb,amsfonts}
\usepackage{amsthm}
\usepackage{mathrsfs}
\usepackage{textcomp}
\usepackage{manyfoot}
\usepackage{booktabs}
\usepackage{listings}
\usepackage{CJKutf8}
\usepackage{mdframed}
\usepackage{tabularx}
\usepackage{caption}
\usepackage{hyperref} 
\usepackage[utf8]{inputenc}
\usepackage[linesnumbered, ruled, vlined]{algorithm2e} 
\usepackage{setspace} 
\usepackage{anyfontsize}

\usepackage{tabularx}
\usepackage{hyperref}

\def\BibTeX{{\rm B\kern-.05em{\sc i\kern-.025em b}\kern-.08em
    T\kern-.1667em\lower.7ex\hbox{E}\kern-.125emX}}
\begin{document}

\title{Optimizing Multi-Stage Language Models for Effective Text Retrieval
}

\author{\IEEEauthorblockN{Quang Hoang Trung*}
\IEEEauthorblockA{\textit{VJ Technology} \\
Da Nang city, VietNam}
\and

\IEEEauthorblockN{Le Trung Hoang}
\IEEEauthorblockA{\textit{VJ Technology} \\
Da Nang city, VietNam}
\and

\IEEEauthorblockN{Nguyen Van Hoang Phuc}
\IEEEauthorblockA{\textit{VJ Technology} \\
Da Nang city, VietNam}
\and

\thanks{(*) Corresponding author(s). E-mail(s): \href{trung.quang@vj-tech.jp}{trung.quang@vj-tech.jp}}%
}

\maketitle

\begin{abstract}
Efficient text retrieval is critical for applications such as legal document analysis, particularly in specialized contexts like Japanese legal systems. Existing retrieval methods often underperform in such domain-specific scenarios, necessitating tailored approaches. In this paper, we introduce a novel two-phase text retrieval pipeline optimized for Japanese legal datasets. Our method leverages advanced language models to achieve state-of-the-art performance, significantly improving retrieval efficiency and accuracy. To further enhance robustness and adaptability, we incorporate an ensemble model that integrates multiple retrieval strategies, resulting in superior outcomes across diverse tasks. Extensive experiments validate the effectiveness of our approach, demonstrating strong performance on both Japanese legal datasets and widely recognized benchmarks like MS-MARCO. Our work establishes new standards for text retrieval in domain-specific and general contexts, providing a comprehensive solution for addressing complex queries in legal and multilingual environments.

\end{abstract}


\section{Introduction}
\begin{figure*}[ht]
    \centering
    \includegraphics[width=1.0\textwidth]{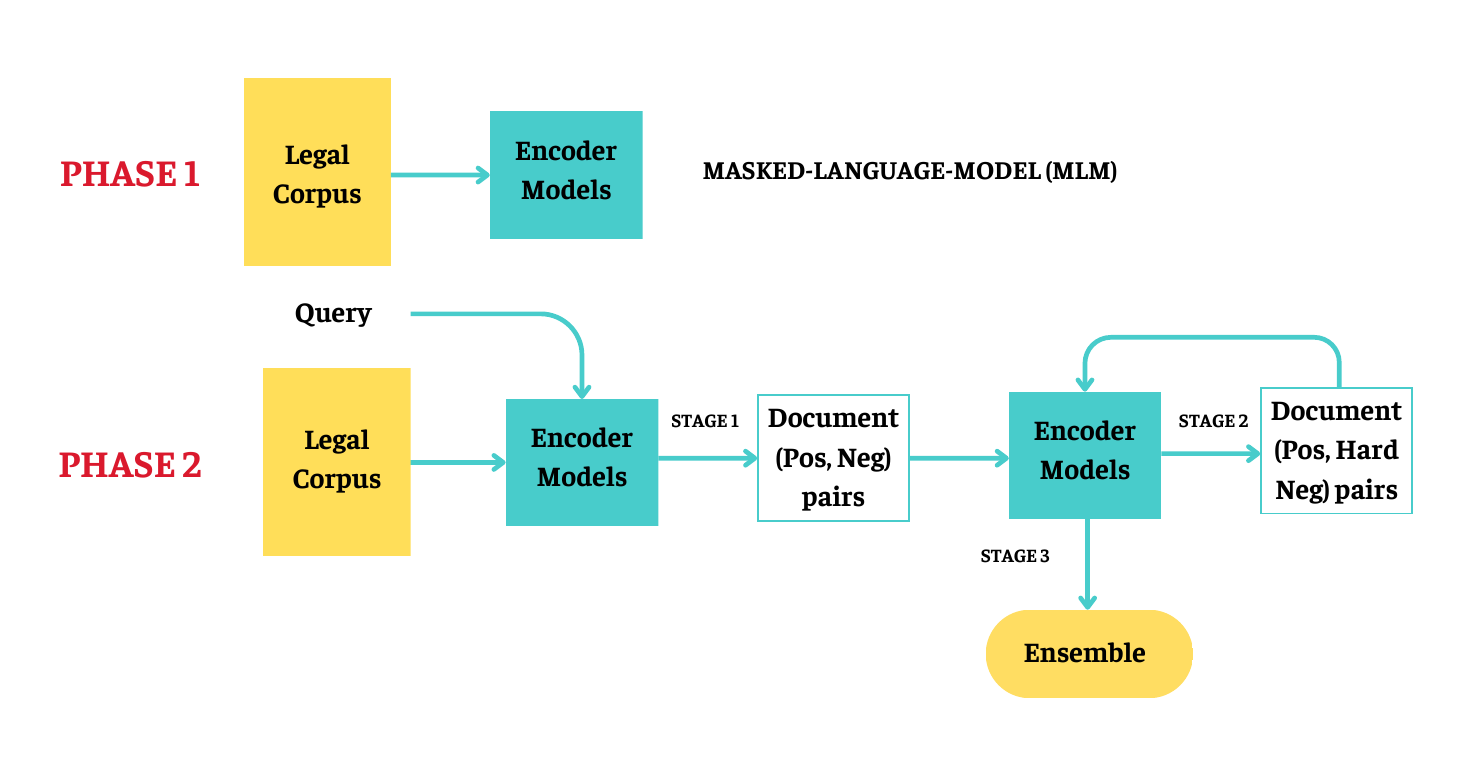}
    \caption{The figure presents an overview of the proposed two-phase text retrieval framework. In Phase 1, the model is pretrained using the Masked Language Model (MLM) task to establish a general contextual understanding of the dataset, creating a strong foundation for subsequent training. Phase 2 consists of three stages. In Stage 1, the encoder model, fine-tuned during Phase 1, is used to retrieve the most relevant documents. Among these, truly relevant documents (labeled as positive) are identified based on human annotations, while documents mistakenly considered relevant are labeled as negative. In Stage 2, both positive and negative documents are input into encoder models, such as BERT or RoBERTa, to further refine the model’s ability to differentiate between relevant and irrelevant documents. Unlike traditional approaches, this method replaces sparse retrieval techniques with language models (LMs) to improve performance. In Stage 3, hard negative examples, generated from the fine-tuned model in Stage 2, are used for additional training to enhance the model’s capacity to address more challenging cases. The process concludes with an ensemble step, combining multiple models or techniques to leverage their individual strengths. This integration minimizes errors, improves accuracy, and enhances the stability of retrieval outcomes, resulting in superior overall performance.}
    \label{fig:image_1}
\end{figure*}

In the field of text retrieval, traditional methods, such as sparse retrieval, have played a foundational role for decades. These approaches, including TF-IDF \cite{salton1988term}, BM25 \cite{robertson2004simple}, and BM25+ \cite{robertson1994some}, rely on the frequency and direct relevance of terms between queries and documents. However, they face significant limitations when addressing semantic matching due to differences in meaning and expression between the query and the document. Despite these challenges, BM25+ remains a strong baseline because of its ability to balance the weight of important keywords and their frequency effectively.

Dense retrieval has significantly advanced text retrieval with the advent of deep learning models and word embeddings. This method encodes both queries and documents into vectors within a continuous semantic space, enabling more accurate semantic comparisons. Recent studies, such as those by \cite{jin2023lader} and \cite{sil2023primeqa}, demonstrate that dense retrieval substantially improves the ability to retrieve relevant documents by alleviating the issue of lexical mismatch.

Within dense retrieval, two widely adopted techniques are the cross-encoder and the bi-encoder. The cross-encoder processes both the query and the document simultaneously, generating a combined representation for enhanced accuracy, though at a high computational cost. Conversely, the bi-encoder independently encodes queries and documents, offering faster processing and greater scalability, particularly when working with large datasets.

More recently, generative retrieval models, such as DSI-QG \cite{zhuang2022bridging} and GENRET \cite{sun2024learning}, have introduced an innovative paradigm. These systems generate document identifiers (docids) or relevant document content directly from the query. By leveraging the power of large language models (LLMs), these approaches enhance retrieval accuracy, especially in scenarios where traditional retrieval methods struggle.

In this study, we introduce a novel Two-phase text retrieval approach specifically optimized for Japanese legal datasets, as derived from the work in paper \cite{trung2024adaptive}

\noindent\textbf{Contributions.} Our contributions are as follows:
\begin{itemize}
\item Development of an innovative retrieval pipeline: Inspired by CoCondenser \cite{gao2021unsupervised}, we propose a new text retrieval method tailored for the Japanese legal context. This approach leverages advanced language models (LMs) to significantly enhance retrieval efficiency and accuracy in Figure \ref{fig:image_1}. The proposed method not only demonstrates superior performance on Japanese legal datasets but also achieves remarkable results on standard benchmarks like MS-MARCO \cite{bajaj2016ms}.
\item Integration of an ensemble model: As detailed in Section \ref{extension}, we expand the pipeline by incorporating an ensemble model that combines multiple retrieval techniques. This integration improves the adaptability and robustness of the system, ensuring high-quality performance across a wide range of retrieval tasks.
\item Comprehensive experimental validation: We conducted extensive experiments to validate the effectiveness of our approach, showcasing its strong performance across diverse scenarios and languages.
\end{itemize}

In summary, this paper introduces a retrieval pipeline specifically designed for Japanese legal contexts, proposes an ensemble extension to enhance retrieval outcomes. Furthermore, our method not only excels in handling domain-specific datasets but also sets new standards of performance on widely recognized benchmarks.


\section{PRELIMINARIES}
\noindent
\textbf{Masked Language Model (MLM)} This is a training technique where certain tokens (words or characters) in a sentence are masked (hidden) by replacing them with a special token, such as \texttt{[MASK]}. The task of the model is to predict the masked tokens based on their context (i.e., the surrounding words).
\\
\\
\textbf{Contrastive Learning} Contrastive learning works by learning a representation that maps similar data points closer in the embedding space and dissimilar data points farther apart. This is achieved by optimizing a \textit{contrastive loss} function, which encourages the model to reduce the distance between similar data points while increasing the distance between dissimilar points. In the context of learning representations for text retrieval, the data points could be queries and documents, with the goal of learning a representation that maps related queries and documents closer together while mapping unrelated ones farther apart.

The general form of the contrastive loss is:

\begin{equation}
L\left( W, \left(Y, \mathbf{X}_1, \mathbf{X}_2\right)^i \right) = (1 - Y)L_S\left(D_W^i\right) + YL_D\left(D_W^i\right)
\label{equal_1}
\end{equation}

Where:
\begin{itemize}
    \item $Y$ determines whether the two data points $(\mathbf{X}_1, \mathbf{X}_2)$ are similar ($Y=0$) or dissimilar ($Y=1$).
    \item $L_S$ represents the loss function for similar data points.
    \item $L_D$ represents the loss function for dissimilar data points.
    \item $D_W$ is a similarity (or dissimilarity) metric between two data points, introduced by LeCun as:
    \begin{equation}
    D_W\left(\mathbf{X}_1, \mathbf{X}_2\right) = \left\| G_W(\mathbf{X}_1) - G_W(\mathbf{X}_2) \right\|_2
    \end{equation}
    Here, $G$ represents the mapping function (typically a neural network). This is an Euclidean distance (L2 norm), but other distance metrics like Manhattan distance, Cosine similarity, etc., can also be used.
\end{itemize}

The formula \ref{equal_1} resembles the Cross-Entropy loss structurally. However, while Cross-Entropy loss operates over class probabilities for classification tasks, the contrastive loss focuses on data point distances in the learned embedding space. This makes contrastive loss particularly suited for tasks like face verification, where the goal is to learn representations without explicitly modeling class distributions.

The exact contrastive loss proposed by LeCun is as follows:


\begin{equation}
L\left( W, \left(Y, \mathbf{X}_1, \mathbf{X}_2\right)^i \right) = 
\begin{aligned}
    & (1 - Y)\frac{1}{2}(D_W)^2 \\
    & + Y\frac{1}{2}\max\{0, m - D_W\}^2
\end{aligned}
\end{equation}

Where:
\begin{itemize}
    \item $L_S$ (loss for similar data points): If two data points are labeled as similar, minimize the Euclidean distance $D_W$ between them.
    \item $L_D$ (loss for dissimilar data points): If two data points are labeled as dissimilar, maximize the Euclidean distance $D_W$ up to a margin $m$.
\end{itemize}

\begin{figure}[t]
    \centering
    \includegraphics[width = 0.9\linewidth]{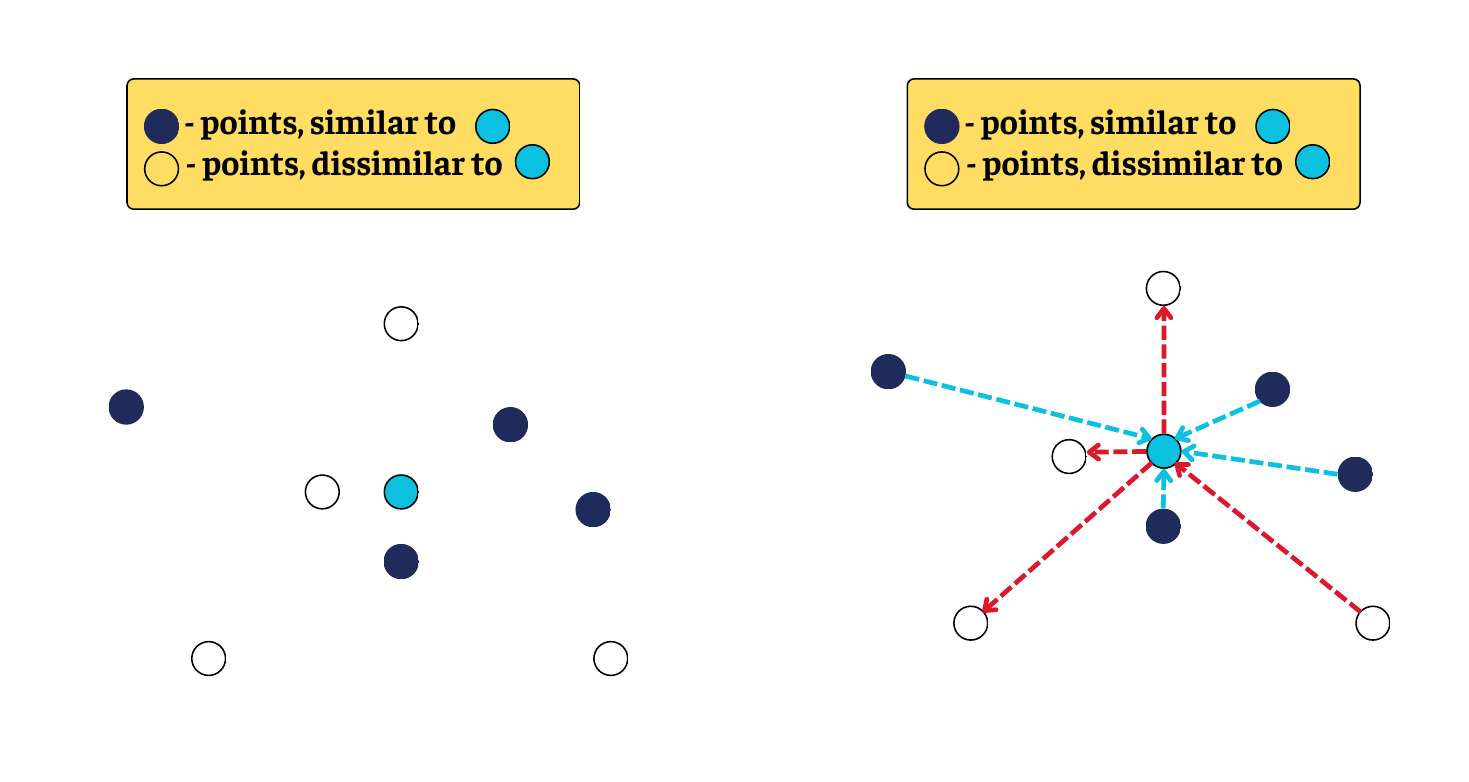}
    	\vspace{-0.2cm}
	\caption{Illustration of contrastive learning - Similar points (black) are grouped closer, dissimilar points (white) are pushed farther.}
    \label{fig: contras_1}
\end{figure}

\begin{figure}[t]
    \centering
    \includegraphics[width = 0.9\linewidth]{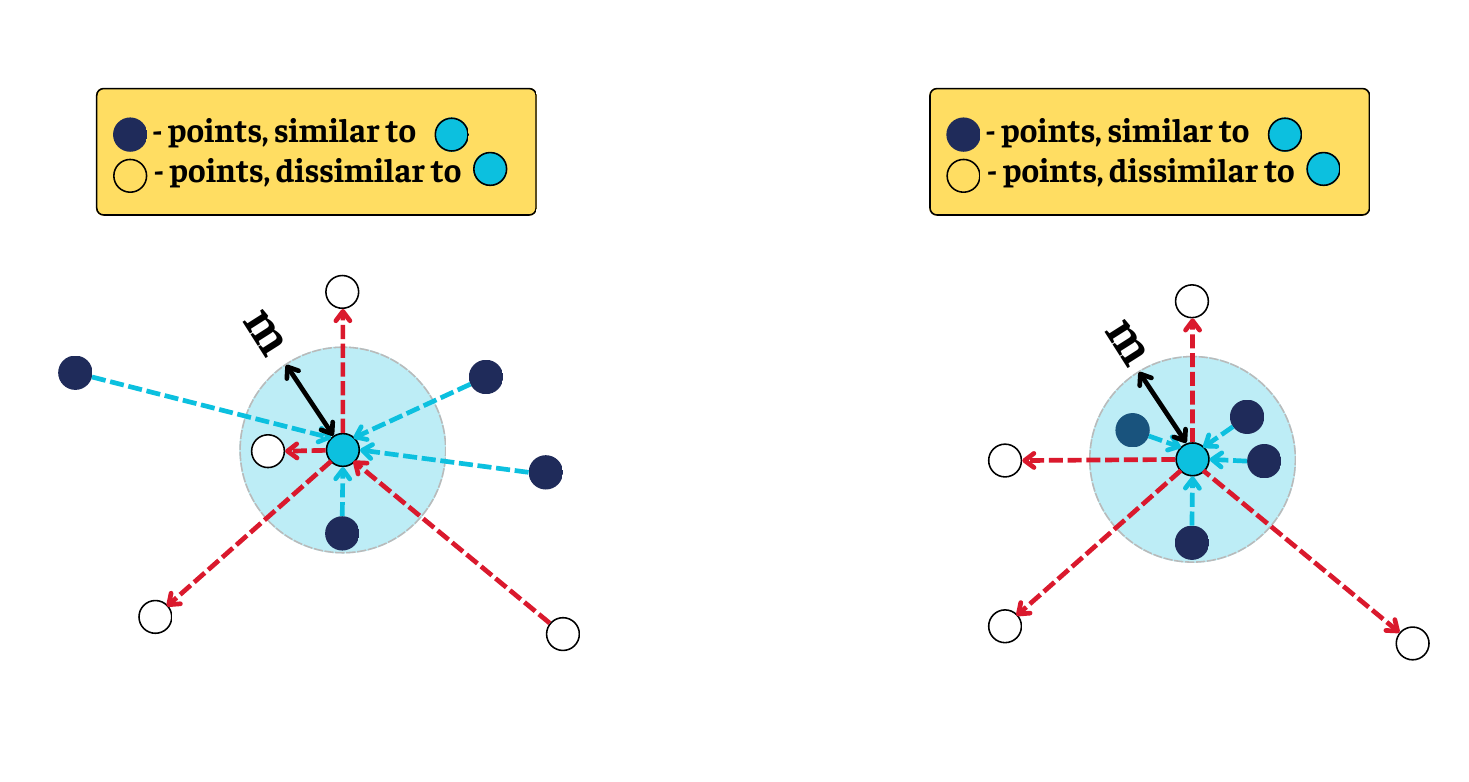}
    	\vspace{-0.2cm}
	\caption{Illustration of decision boundaries using contrastive loss.}
    \label{fig: contras_2}
\end{figure}

The goal of contrastive learning is to ensure that for each class/group of similar data points, the intra-class distance is minimized while the inter-class distance is maximized. This is illustrated in Figures \ref{fig: contras_1} and \ref{fig: contras_2}.

For any given data point, we ensure that:
\begin{itemize}
    \item White points (dissimilar) lie outside the boundary defined by margin $m$.
    \item Black points (similar) lie within the boundary.
\end{itemize}

When adding a new data point, we use the nearest-neighbor algorithm to determine its similarity/dissimilarity based on whether it lies within the margin $m$.
\\

\begin{figure}[t]
    \centering
    \includegraphics[width = 0.9\linewidth]{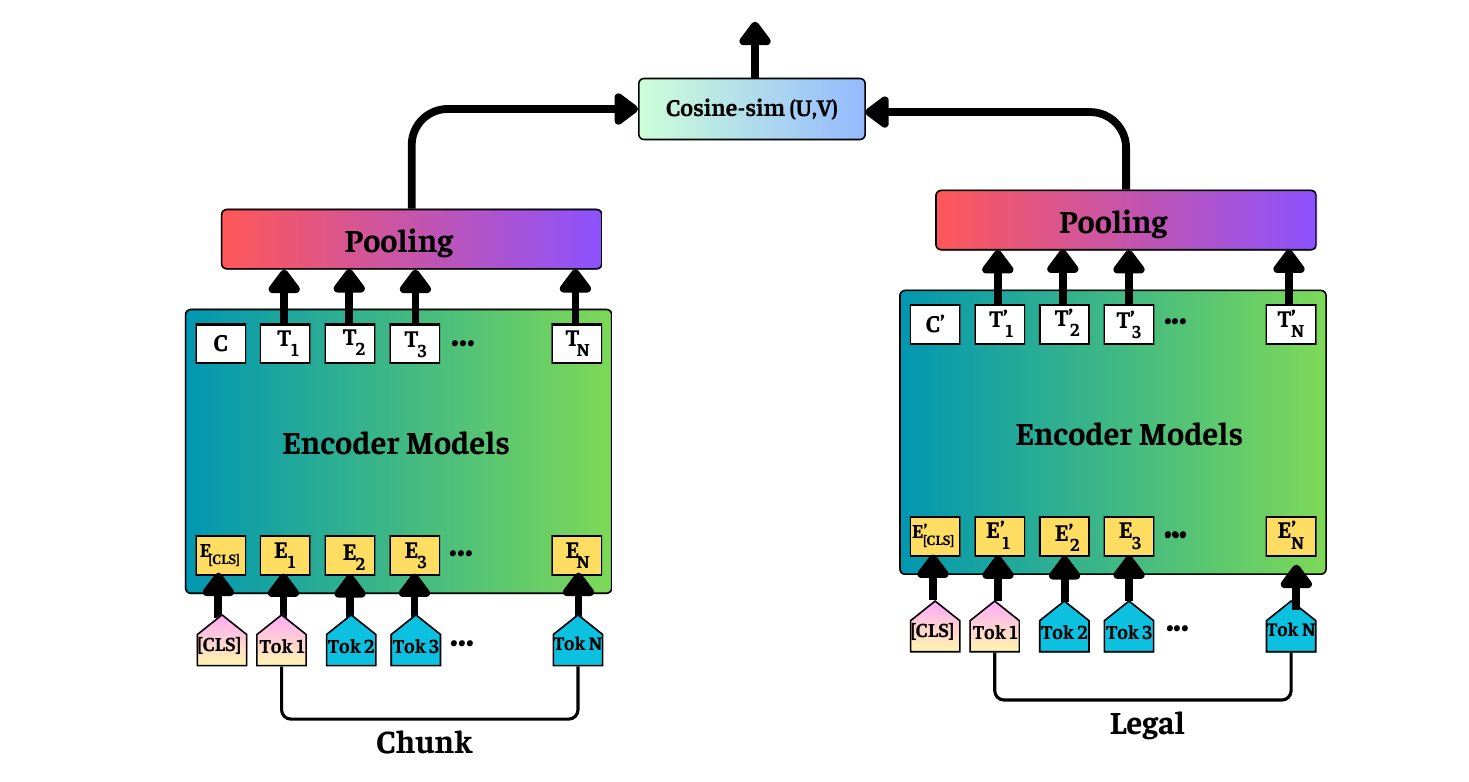}
    	\vspace{-0.2cm}
	\caption{Illustration of sentence similarity calculation, dual encoder models process text chunks and legal passages, applying pooling and cosine similarity for semantic matching.}
    \label{fig: fig5}
\end{figure}

The contrastive loss ensures that:
\begin{itemize}
    \item If $D_W \geq m$ for dissimilar data points, the penalty becomes zero since $\max\{0, m - D_W\} = 0$. This avoids unnecessarily pushing dissimilar points farther than required.
    \item For similar data points, $D_W$ is minimized, ensuring that they are closer in the embedding space.
\end{itemize}
\noindent
\textbf{Utilize contrastive loss to train the encoder model} As shown in Figure \ref{fig: fig5}, the input consists of a query (chunk) and documents (legal), which are processed independently by two separate encoder models. The outputs are pooled to generate two vectors representing the query and the documents. These vectors are then compared using cosine similarity, and the process is optimized with contrastive loss, as detailed in Algorithm \ref{algo:contrastive_loss}.

\begin{algorithm}[t]

\footnotesize 
\caption{Contrastive Loss with Cosine Similarity}\label{algo:contrastive_loss}
\KwIn{$chunk$, $legal$, $model$, margin $m = 0.5$, similarity $score$}
\KwOut{$loss$}

\textbf{Step 1: Generate vector embeddings:} \\
\Indp 
$chunk\_embedding \gets model.embedding(chunk)$ \\
$legal\_embedding \gets model.embedding(legal)$ \\
\Indm 

\textbf{Step 2: Compute cosine similarity:} \\
\Indp
$distances \gets cosine\_similarity(chunk\_embedding, legal\_embedding)$ \\
\Indm

\textbf{Step 3: Define the margin:} \\
\Indp
$m \gets 0.5$ \\
\Indm

\textbf{Step 4: Calculate loss function:} \\
\Indp
$loss \gets 0.5 \cdot \Big(score \cdot distances^2 + (1 - score) \cdot ReLU(m - distances)^2\Big)$ \\
\Indm

\Return $loss$ \\
\end{algorithm}

\section{PROPOSED APPROACH}
In this paper, we introduce proposed pipelines for text retrieval, relying entirely on Language Models (LMs) and eliminating the need for sparse retrieval techniques such as BM25 or BM25+. Unlike the CoCondenser \cite{gao2021unsupervised}, which combines BM25+ with pre-trained language models, our approach replaces BM25+ entirely with language models to achieve superior performance. The proposed methods are described as follows:

\subsection*{\textbf{LMS (Round 1):}~Pretraining and Fine-tuning in Two Phases}

The first pipeline consists of two distinct phases:
\begin{itemize}
    \item Phase 1: The encoder model is initialized with pre-trained weights and further enhanced using a Masked Language Model (MLM) objective. This step helps the encoder better understand contextual relationships within the corpus.\\ \\
    \item Phase 2: 
    \begin{itemize}
        \item Stage 1: Using the pretrained language model from Phase 1, the system generates document pairs (positive and negative) based on the query and the Legal Corpus.
        \item Stage 2: The encoder model, fine-tuned on data generated in Stage 1, is used to distinguish between relevant (positive) and irrelevant (negative) documents.
    \end{itemize}
\end{itemize}

\subsection*{\textbf{LMS (Round 2):}~Refinement with Stage 3 for Hard Negatives}

This pipeline builds upon LMS (Round 1) by introducing an additional stage:
\begin{itemize}
    \item \textbf{Stage 3:} The encoder model is further refined by re-training on hard negative document pairs generated in Stage 2. This step enhances the encoder’s ability to handle difficult or borderline cases, resulting in improved document discrimination.
\end{itemize}

\subsection*{\textbf{LMS (Finetuned MLM) Round 1:}~Leveraging Finetuned MLM for Enhanced Retrieval}
In this pipeline, the Masked Language Model (MLM) fine-tuned in Phase 1 is directly integrated into the retrieval process:

\begin{itemize}
    \item Phase 2:
    \begin{itemize}
        \item Stage 1: The encoder, finetuned with the MLM objective in Phase 1, generates top-ranked documents with the highest relevance to the query.
        \item Stage 2: The encoder model, further trained on contrastive loss using the data generated in Stage 1, distinguishes relevant documents more effectively.
    \end{itemize}
\end{itemize}

\subsection*{\textbf{LMS (Finetuned MLM) Round 2:}~Refining Finetuned Models for Greater Precision}

This pipeline extends LMS (Finetuned MLM) Round 1 by incorporating Stage 3:
\begin{itemize}
    \item Stage 3: Similar to LMS (Round 2), the encoder is re-trained on hard negative document pairs generated in Stage 2. This additional training step sharpens the encoder’s ability to distinguish between closely similar and dissimilar documents, achieving the best possible performance.
\end{itemize}

These pipelines collectively demonstrate the effectiveness of fully replacing sparse retrieval methods with Language Models for superior text retrieval performance. The iterative refinement through stages enables the model to handle increasingly challenging data distributions while maintaining contextual relevance.
\section{Extensions}
\label{extension}
\begin{table*}[!t]
\caption{A wide range of weight combinations derived from grid search for optimizing ensemble performance on Japanese Legal Dataset.}
\label{tab:evaluation_results}
\centering
\small 
\begin{tabularx}{\textwidth}{ccc|XXXXXXXXXXX}
\toprule
\textbf{$\alpha$} & \textbf{$\beta$} & \textbf{$\theta$} & \textbf{R@3} & \textbf{R@5} & \textbf{R@10} & \textbf{R@20} & \textbf{R@50} & \textbf{R@100} & \textbf{R@200} & \textbf{MAP@10} & \textbf{MRR@10} & \textbf{nDCG@10} \\
\midrule
0.3 & 0.25 & 0.45 & \textbf{0.72} & 0.77 & \textbf{0.85} & \textbf{0.87} & \textbf{0.93} & 0.94& \textbf{0.99} & \textbf{0.71} & \textbf{0.91} & \textbf{0.78} \\
0.4 & 0.3 & 0.3 & 0.68 & 0.74 & 0.84 & 0.86 & 0.91 & 0.95 & 0.98 & 0.66 & 0.86 & 0.75 \\
0.2 & 0.5 & 0.3 & 0.68 & 0.73 & 0.82 & 0.86 & 0.90 & 0.95 & 0.98 & 0.66 & 0.87 & 0.74 \\
0.5 & 0.2 & 0.3 & 0.68 & 0.73 & 0.83 & 0.86 & 0.91 & 0.95 & 0.98 & 0.65 & 0.85 & 0.74 \\
0.25 & 0.35 & 0.4 & 0.71 & \textbf{0.78} & 0.84 & 0.86 & 0.91 & 0.95 & 0.98 & 0.69 & 0.89 & 0.77 \\
0.3 & 0.3 & 0.4 & 0.71 & 0.77 & 0.84 & 0.86 & 0.90 & 0.95 & 0.98 & 0.69 & 0.89 & 0.77 \\
0.2 & 0.3 & 0.5 & 0.71 & 0.76 & 0.84 & 0.85 & 0.91 & \textbf{0.96} & 0.98 & 0.70 & 0.88 & 0.77 \\
0.4 & 0.4 & 0.2 & 0.63 & 0.67 & 0.79 & 0.85 & 0.91 & 0.95 & 0.96 & 0.61 & 0.82 & 0.70 \\
\bottomrule
\end{tabularx}
\end{table*}

\begin{figure}[t]
    \centering
    \includegraphics[width = 0.9\linewidth]{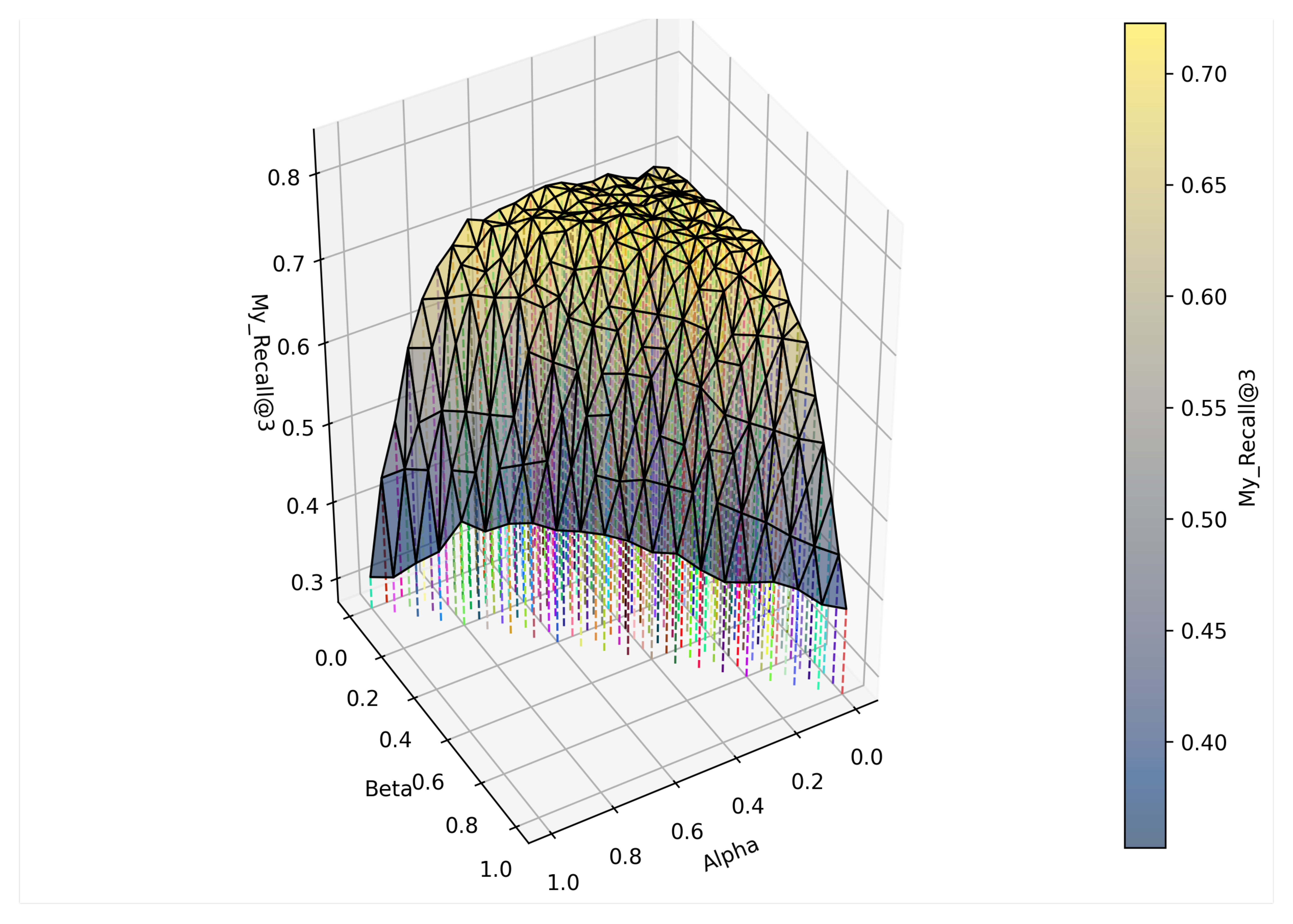}
    	\vspace{-0.2cm}
	\caption{Visualization of the grid search process, depicting My\_Recall@3 scores across different weight combinations.}
    \label{fig: extension_fig}
\end{figure}

\noindent
\textbf{Ensemble Methodology} To further enhance the performance of our proposed retrieval framework, we adopt an ensemble approach that combines the outputs of three models: \textit{\textbf{paraphrase-multilingual-mpnet-base-v2}}, \textit{\textbf{distiluse-base-multilingual-cased-v1}}, and the \textit{\textbf{LMS (Finetuned MLM) Round 2}}. Each model captures unique aspects of the retrieval task, and combining their outputs enables the framework to leverage their complementary strengths. The ensemble score is computed using the formula:
\begin{equation}
\text{Score} = \alpha \cdot \text{score}_{\text{model1}} + \beta \cdot \text{score}_{\text{model2}} + \theta \cdot \text{score}_{\text{model3}}
\end{equation}
\\
where $\alpha$ +  $\beta$ + $\theta$ and the score represents the $My\_Recall@3$ metric used for evaluation in Figure \ref{fig: extension_fig}. The coefficients $\alpha$, $\beta$ and $\theta$ serve as the weights assigned to the contributions of each model.
\\
\\
\textbf{Weight Optimization Process} To identify the optimal set of weights, we performed a systematic grid search across possible values of $\alpha$ and $\beta$, adhering to the constraint $\alpha + \beta + \theta = 1$. The search process involved:
\begin{enumerate}
    \item Varying $\alpha$ incrementally from 0 to 1 in steps of 0.05.
    \item For each value of $\alpha$, varying $\beta$ from 0 to $1 - \alpha$ in steps of 0.05.
    \item Calculating $\theta$ as $\theta = 1 - (\alpha + \beta)$ to satisfy the total weight constraint.
\end{enumerate}

This process generated a comprehensive set of valid weight combinations, systematically covering the space of possible distributions. Each combination was evaluated based on the ensemble's My\_Recall@3 performance.
\\
\\
\textbf{Results and Observations} The grid search identified the optimal weight combination as $\alpha = 0.3$, $\beta = 0.25$, and $\theta = 0.45$, resulting in a final My\_Recall@3 score of 0.72. This configuration demonstrates the effectiveness of the ensemble approach, significantly improving performance over individual models. The relatively larger weight assigned to the \textit{\textbf{LMS (Finetuned MLM) Round 2}} ($\theta = 0.45$) highlights its critical role in the ensemble, benefiting from task-specific tuning.

Figure \ref{fig: extension_fig} visualizes the grid search process, showing the progression of My\_Recall@3 scores across various weight combinations. The figure highlights regions of high performance and illustrates how the ensemble's effectiveness depends on the weight distribution.
\\
\\
\textbf{Example Weight Combinations} Table \ref{tab:evaluation_results} presents sample weight combinations from the grid search, including the optimal set, to illustrate the systematic evaluation process. This demonstrates the robustness of the ensemble method, as even suboptimal configurations yield competitive My\_Recall@3 scores.

\section{Experiments}

\begin{table*}[!t]
\centering
\caption{Performance of our method on the Japanese Legal Dataset, highlighting significant advancements in retrieval effectiveness over baseline approaches and existing techniques.}\label{tab:ja_eval}
\resizebox{\textwidth}{!}{
\begin{tabular}{lccccccc}
\toprule
\textbf{Method} & \textbf{R@3} & \textbf{R@5} & \textbf{R@10} & \textbf{R@20} & \textbf{R@50} & \textbf{R@100} & \textbf{R@200} \\
\midrule
\multicolumn{8}{l}{\textit{Sparse Retrieval}} \\
TF-IDF \cite{salton1988term} & 42.82 & 46.18 & 55.11 & 63.19 & 73.29 & 78.07 & 83.76 \\
BM25+ \cite{robertson1994some} & 37.82 & 41.09 & 50.59 & 57.1 & 65.84 & 73.51 & 81.45 \\
\midrule
\multicolumn{8}{l}{\textit{Dense Retrieval}} \\
Siamese \cite{reimers2019sentence} & 46.79 & 53.56 & 62.16 & 67.23 & 80.66 & 86.64 & 92.08 \\
Two-Towers Siamese \cite{yang2019multilingual} & 47.82 & 55.63 & 62.23 & 70.12 & 80.29 & 85.74 & 91.99 \\
GC-DPR \cite{karpukhin2020dense} & 42.56 & 46.42 & 53.05 & 63.10 & 76.18 & 82.49 & 87.96 \\
CoCondenser \cite{gao2021unsupervised} & 60.77 & 62.02 & 70.89 & 77.10 & 82.26 & 87.34 & 92.92 \\
3SRM \cite{sasazawa2023text} & 39.87 & 42.38 & 50.13 & 59.95 & 76.85 & 86.76 & 90.08 \\
BM25Plus (Round 2) \textbf{(Ours)} & 65.38 & 68.33 & 75.91 & 78.93 & \textbf{87.99} & 92.36 & 96.71 \\
LMS (Round 1) \textbf{(Ours)} & 62.95 & 67.90 & 73.52 & 76.81 & 86.47 & 90.40 & 96.60 \\
LMS (Round 2) \textbf{(Ours)} & \textbf{69.10} & 72.39 & 78.73 & 80.27 & 87.48 & 92.33 & 94.23 \\
LMS (Finetuned MLM) Round 1 \textbf{(Ours)} & 64.87 & 67.32 & 74.91 & 78.02 & 86.86 & \textbf{94.03} & 96.73 \\
LMS (Finetuned MLM) Round 2 \textbf{(Ours)} & 68.72 & \textbf{72.50} & \textbf{79.53} & \textbf{82.57} & 85.67 & 91.58 & \textbf{97.46} \\
\midrule
\multicolumn{8}{l}{\textit{Generative Retrieval}} \\
DSI \cite{tay2022transformer} & 66.03 & 68.86 & 75.21 & 80.34 & 86.38 & 89.19 & 92.90 \\
DSI-QG \cite{zhuang2022bridging} & 15.75 & 23.29 & 28.77 & 39.49 & 56.85 & 68.72 & 84.24 \\
\bottomrule
\end{tabular}
}
\end{table*}

To evaluate the effectiveness of our proposed retrieval pipeline, we conducted experiments on Japanese legal datasets derived from the work in \cite{trung2024adaptive} as well as a split of the MS MARCO passage dataset \cite{bajaj2016ms}. Our method, inspired by CoCondenser \cite{gao2021unsupervised}, leverages advanced language models to enhance retrieval efficiency and accuracy, particularly within the Japanese legal domain. These experiments validate the robustness of our approach, demonstrating its strong performance on both domain-specific and standard benchmarks.
\\
\\
\subsection{Legal Japanese Retrieval Dataset}
\noindent
\textbf{Dataset} Based on the work in \cite{trung2024adaptive}, this dataset is specifically tailored to the Japanese legal context. It comprises 3,259 training examples, 73 validation examples, and 130 test examples. Key evaluation metrics include Recall, MRR@10, MAP@10, and nDCG@10, which assess retrieval performance within this domain-specific dataset.
\\
\\
\textbf{In-Domain Evaluation} Table \ref{tab:ja_eval} highlights the superior performance of our proposed methods compared to sparse, dense, and generative retrieval approaches on Japanese legal datasets 

Sparse Retrieval, such as TF-IDF and BM25+ rely on keyword frequency-based matching. These methods often struggle to capture the semantic relationships necessary for effective retrieval in complex datasets. Consequently, their performance is suboptimal across all metrics. For example, at Recall@200, TF-IDF achieves 83.76, while BM25+ lags behind at 81.45. By contrast, \textit{\textbf{LMS (Finetuned MLM) Round 2}} achieves a significantly higher Recall@200 of 97.46, surpassing TF-IDF and BM25+ by 13.7 and 16.01 points, respectively.

This disparity becomes even more pronounced at lower recall levels. At Recall@3, TF-IDF and BM25+ achieve 42.82 and 37.82, respectively, whereas our method \textit{\textbf{BM25Plus (Round 2)}} reaches 65.38, outperforming TF-IDF by 22.56 points and BM25+ by 27.56 points. These results illustrate the fundamental limitations of sparse retrieval in handling nuanced queries and emphasize the importance of leveraging language models for semantic understanding.

Dense Retrieval leverage language models (LMs) to encode semantic representations of queries and documents. This category includes models like Siamese \cite{reimers2019sentence}, Two-Towers Siamese \cite{yang2019multilingual}, GC-DPR \cite{karpukhin2020dense}, and CoCondenser \cite{gao2021unsupervised}. Although dense methods outperform sparse retrieval approaches, they are consistently outperformed by our methods.

For example, at Recall@10, CoCondenser achieves 70.89, outperforming other dense retrieval baselines such as Siamese (62.16) and GC-DPR (53.05). However, \textit{\textbf{LMS (Round 2)}} further improves on CoCondenser with a Recall@10 of 78.73, demonstrating a 7.84-point lead. This advantage persists across higher recall levels, with our method achieving 94.23 at Recall@200 compared to CoCondenser’s 92.92, marking a 1.31-point improvement.

The improvements can be attributed to our two-phase training pipeline, particularly Phase 2 Stage 3, where hard negative examples are incorporated into the training process. This refinement allows our methods to address challenging cases and ambiguous document-query relationships, which dense retrieval methods often fail to resolve.

Generative Retrieval, such as DSI \cite{tay2022transformer} and DSI-QG \cite{zhuang2022bridging}, attempt to directly model query-document relationships by generating document representations. While DSI achieves competitive results at lower recall levels (e.g., Recall@3 of 66.03), its performance declines for higher recall levels. At Recall@200, for instance, DSI achieves 92.90, falling behind our best method \textit{\textbf{LMS (Finetuned MLM) Round 2}} by 4.56 points (97.46 vs. 92.90).

\begin{table*}[!t]
\centering
\caption{Evaluation of our method on the MS MARCO Benchmark, showcasing substantial improvements across retrieval metrics compared to baseline approaches and existing methods.}
\label{tab:benchmark}
\resizebox{\textwidth}{!}{
\begin{tabular}{lcccccc}
\toprule
\textbf{Model/Metrics} & \textbf{R@3} & \textbf{R@100} & \textbf{MRR@10} & \textbf{MAP@10} & \textbf{nDCG@10} \\
\midrule
\multicolumn{6}{l}{\textit{Sparse Retrieval}} \\
TF-IDF & 26.9 & 83.6 & 43.8 & 23.2 & 30.6 \\
BM25+ & 28.9 & 73.5 & \textbf{49.3} & 24.5 & 30.9 \\
\midrule
\multicolumn{6}{l}{\textit{Dense Retrieval}} \\
BM25Plus (Round 1) & 29.2 & 73.5 & 25.9 & 24.6 & 30.0 \\
BM25Plus (Round 2) \textbf{(Ours)} & \textbf{34.5} & \textbf{84.1} & 30.2 & \textbf{29.0} & \textbf{36.1} \\
\midrule
\multicolumn{6}{l}{\textit{Generative Retrieval}} \\
DSI & 14.3 & 40.9 & 13.7 & 13.6 & 15.5 \\
\bottomrule
\end{tabular}
}
\end{table*}

DSI-QG, which incorporates query generation to address documents without associated queries, struggles significantly with the linguistic complexities of Japanese and produces suboptimal results. It achieves only 28.77 at Recall@10, far below even sparse retrieval methods like BM25+ (50.59) and our methods, such as \textit{\textbf{LMS (Round 1)}}, which achieves 73.52. This gap highlights the challenges generative methods face in creating high-quality semantic representations for non-English datasets.

Our proposed methods consistently outperform baselines across all metrics. Among them, \textit{\textbf{LMS (Finetuned MLM) Round 2}} achieves the highest performance, setting new benchmarks in Recall@5 (72.50), Recall@10 (79.53), and Recall@200 (97.46).
\begin{itemize}
\item Comparison with Sparse Retrieval: Compared to BM25+, our method improves Recall@20 by 25.47 points (82.57 vs. 57.10) and Recall@200 by 16.01 points (97.46 vs. 81.45).
\item Comparison with Dense Retrieval: At Recall@50, our method achieves a 3.41-point improvement over CoCondenser (85.67 vs. 82.26) and a 5.38-point improvement over Two-Towers Siamese (85.67 vs. 80.29).
\item Comparison with Generative Retrieval: Our method achieves significantly higher Recall@5, Recall@10 compared to DSI (72.50 vs. 68.86 and 79.53 vs. 75.21, respectively).
\end{itemize}

Effect of Multi-Stage Training: The superior performance of our methods can be attributed to Phase 2 Stage 3, where fine-tuned models are trained on hard negative examples. For example, \textit{\textbf{LMS (Finetuned MLM) Round 1}}, which does not include Stage 3, achieves Recall@10 of 74.91, while \textit{\textbf{LMS (Finetuned MLM) Round 2}} achieves 79.53, demonstrating a 4.62-point improvement.

Summary: The results underscore the robustness of our proposed retrieval pipeline. By replacing traditional sparse retrieval techniques with advanced language models and incorporating multi-stage training, our methods achieve state-of-the-art performance across all metrics. These improvements highlight the scalability and effectiveness of our approach, especially in the Japanese legal domain.

\subsection{Benchmark Datasets Retrieval}
\noindent
\textbf{Dataset} We utilized a split of the MS MARCO passage dataset, which originally includes around 1.01 million rows. For this study, we employed a subset of 17,132 rows, divided into 15,270 for training, 1,000 for validation, and 862 for testing. The corpus contained approximately 134,000 documents. This split was designed to ensure a manageable yet representative subset for experimentation. Stage 2 training involved regular negatives, while Stage 3 focused on hard negatives, as detailed in the experimental setup.
\\
\\
\textbf{In-Domain Evaluation} Table \ref{tab:benchmark} illustrates the effectiveness of our proposed retrieval methods across various evaluation metrics on the training split of the MS MARCO passage dataset. The results underscore the superiority of our approaches, \textit{\textbf{BM25Plus (Round 1)}} and \textit{\textbf{BM25Plus (Round 2)}}, over baseline sparse and generative retrieval techniques.

Sparse retrieval perform moderately well but are inherently constrained by their reliance on exact lexical matches. For instance, BM25+ achieves a Recall@3 of 28.9 and MAP@10 of 24.5, while its nDCG@10 is higher at 30.9, indicating that its ranking quality benefits from a focus on top results. Similarly, TF-IDF demonstrates strong performance in Recall@100 with a value of 83.6 but lags in Recall@3 (26.9) and MAP@10 (23.2). Notably, BM25+ achieves the highest MRR@10 among sparse methods at 49.3, surpassing our proposed methods in this metric. This suggests that BM25+'s lexical matching mechanisms can still be effective in identifying highly relevant documents for ranking. However, these methods struggle to fully capture semantic nuances, which limits their ability to perform consistently across a range of retrieval metrics in complex scenarios.

Dense Retrieval, in contrast, our proposed methods leverage dense retrieval techniques to overcome these limitations and deliver significant improvements across most metrics. \textit{\textbf{BM25Plus (Round 1)}}, which omits hard negative training in Phase 2 Stage 3, achieves a Recall@3 of 29.2 and MAP@10 of 24.6, closely rivaling BM25+’s performance. Notably, when hard negative training is incorporated, \textit{\textbf{BM25Plus (Round 2)}} outperforms BM25+ in Recall@3 (34.5 vs. 28.9), MAP@10 (29.0 vs. 24.5), and nDCG@10 (36.1 vs. 30.9). These results highlight the robustness of our methods, particularly in metrics focused on ranking quality.

Generative retrieval perform poorly across all metrics, achieving only a Recall@3 of 14.3 and nDCG@10 of 15.5. These results emphasize the inherent limitations of generative approaches in retrieval tasks. The reliance on generative language modeling often lacks the precision and relevance required for effective ranking. For instance, when compared to \textit{\textbf{BM25Plus (Round 2)}}, which achieves a Recall@3 of 34.5 and nDCG@10 of 36.1, DSI’s performance falls significantly short. This highlights its inability to handle ranking tasks involving complex semantic nuances.

Furthermore, DSI’s lack of capability to leverage explicit document-query interactions is a key limitation that constrains its effectiveness. In contrast, \textit{\textbf{BM25Plus (Round 2)}} utilizes iterative training with hard negatives, refining ranking capabilities to achieve a MAP@10 of 29.0, far exceeding DSI’s 13.6. These metrics underscore the robustness and adaptability of our approach in addressing nuanced relationships required for high-quality document retrieval. This makes it a clearly superior alternative to generative paradigms.

Proposed Methods: The two-phase training strategy employed in our proposed methods underpins their superior performance. In Phase 1, pretraining with the Masked Language Model (MLM) task establishes a strong foundational understanding of the dataset. Phase 2’s multi-stage approach, which involves the initial BM25+, refinement with positive and negative samples using language models like XLM-RoBERTa, and iterative training with hard negatives, ensures a comprehensive and nuanced model training process. This progression enables our methods to outperform traditional sparse retrieval techniques and state-of-the-art generative retrieval models consistently.

Overall, these findings emphasize the robustness and adaptability of our retrieval framework, demonstrating its capability to excel across a diverse range of retrieval metrics. The significant performance gains achieved by \textit{\textbf{BM25Plus (Round 2)}}  highlight the value of incorporating advanced training strategies and iterative refinement in modern retrieval tasks.

\section{Conclusion}

This paper presents a novel approach to Japanese legal text retrieval, introducing a tailored two-phase retrieval pipeline and an ensemble model to enhance retrieval performance. By leveraging advanced language models, our method achieves state-of-the-art results, excelling in both domain-specific datasets and widely recognized benchmarks. The proposed pipeline addresses the unique challenges of the Japanese legal context, providing a robust and adaptable solution for complex retrieval tasks.

Empirical evaluations demonstrate the effectiveness of our approach, validating its applicability across diverse scenarios and datasets. Additionally, the integration of an ensemble model ensures consistent performance improvements, further establishing the versatility of the proposed framework.

\bibliographystyle{apalike}
\bibliography{sn-bibliography}


\end{document}